\documentclass[prl,twocolumn,english,showpacs,nofootinbib]{revtex4-2}
\usepackage{amssymb}
\usepackage{color}
\usepackage{graphicx}
\usepackage{bbold}
\usepackage{bm}
\usepackage{amsmath}
\usepackage{amsfonts}
\usepackage{amssymb}
\usepackage{braket}
\usepackage{units}
\usepackage{upgreek}
\usepackage{xcolor}
\usepackage{fancyhdr}
\usepackage{float}
\usepackage{textcmds}
\definecolor{burgundy}{RGB}{97,0,35}
\definecolor{marine}{RGB}{4,46,96}
\definecolor{greendark}{RGB}{3,53,0}
\usepackage{hyperref}
\hypersetup{%
    colorlinks=true,
    urlcolor=marine,
    linkcolor=marine,
    citecolor=marine
    }

\usepackage[normalem]{ulem}
\usepackage{lipsum}

\usepackage{orcidlink}

\begin{document}

	\title{Thermodynamics and density fluctuations in a bilayer Hubbard system of ultracold atoms}

	\author{J. Samland\orcidlink{0009-0001-9326-3402}, N. Wurz, M. Gall and M. K\"ohl}
	\affiliation{Physikalisches Institut, University of Bonn, Wegelerstra{\ss}e 8, 53115 Bonn, Germany}

\begin{abstract}
We measure the equation of state in a bilayer Hubbard system for different ratios of the two tunnelling amplitudes $t_\perp /t$. From the equation of state we deduce the compressibility and observe its dependency on $t_\perp /t$. Moreover, we infer thermodynamic number fluctuations from the equation of state by employing the fluctuation-dissipation theorem. By comparing the thermodynamic with local density fluctuations, we find that non-local density fluctuations in our bilayer Hubbard system become more prominent for higher $t_\perp /t$ in the low filling regime. To validate our measurements, we compare them to Determinant Quantum Monte Carlo simulations of a bilayer Hubbard system with 6$\times$6 lattice sites per layer.
\end{abstract}

\maketitle


The interplay between kinetic energy, interaction strength, and dimensionality in strongly correlated, fermionic quantum systems determines the characteristics of quantum phases. There are numerous examples of collective behaviour emerging in many-body systems. For example, in periodic lattice potentials with an average filling of half a spin-up fermion and half a spin-down fermion per lattice site, strong repulsive interactions induce a Mott-insulating phase in which each lattice site is occupied by precisely one fermion. This microscopic behaviour goes along with the macroscopic property of a vanishing isothermal compressibility. In contrast, for weak interactions the system is in a metallic phase with finite compressibility and delocalised particles.

The fermionic Hubbard model is a simple model for interacting, fermionic particles in a lattice and has been intensively examined over the last years. Here, analogue quantum simulators with cold atoms in optical lattices have proven to be an insightful platform for studying this model and experimentally verifying advanced theoretical models. This had led for example to the observation of the cross-over from a metallic to a Mott insulating phase for increasing repulsive interactions \cite{Jordens2008,Schneider2008,cheuk_observation_2016-1,greif_site-resolved_2016,cocchi_equation_2016} and anti-ferromagnetic ordering due to the super-exchange \cite{Greif_2013, Hart_2015, Cheuk_2016, cheuk_observation_2016-1, greif_site-resolved_2016, drewes_antiferromagnetic_2017}.  Moreover, the behaviour of the  microscopic correlations has been related to thermodynamic properties such as the compressibility \cite{Hofrichter_2016,drewes_thermodynamics_2016,cocchi_measuring_2017-2}.

So far, most experiments regarding the in-situ detection of correlations were conducted in two-dimensional samples. However, recently, also bilayer Hubbard models were implemented, which allow for investigating the effects of controllably coupling two two-dimensional layers to one another \cite{koepsell_robust_2020,gall_competing_2021}. It was shown, for example, that the inter-layer coupling is essential for understanding the magnetic correlations: controlling the ratio between the inter-layer tunneling rate, $t_\perp$, and the in-plane tunneling rate $t$  induces a cross-over from an intra-plane anti-ferromagnetically (AFM) ordered phase to a quantum phase with inter-layer AFM correlations \cite{Golor2014,scalettar_magnetic_1994,kancharla_band_2007,bouadim_magnetic_2008,ruger_phase_2014,Scalettar1995,gall_competing_2021} or can enhance pairing \cite{Bohrdt2022}.

\begin{figure*}
	\centering
	\includegraphics[width=0.75\textwidth]{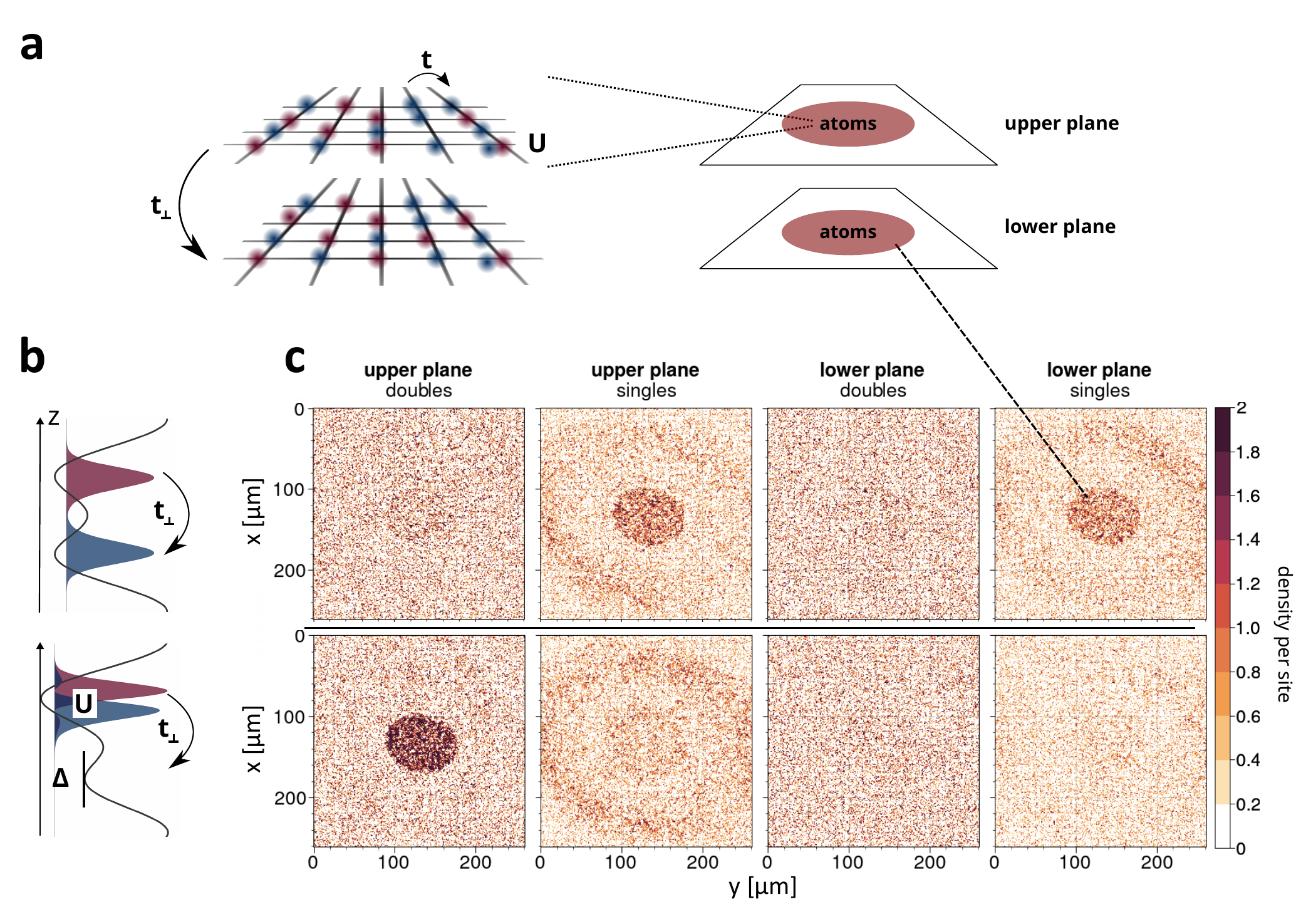}
	\caption{\textbf{a} Schematic of the bilayer Hubbard model  realized in our experiment. The inter-layer tunneling rate $t_\perp$ and the in-plane tunneling rate $t$ are independently tuneable. We image the two layers separately by combining high-resolution microscopy with tomographic addressing of the individual layers. \textbf{b} Visualization of the balanced ($\Delta=0$) and imbalanced ($\Delta \neq 0$) double-well potentials. The energy offset $\Delta$ determines the population of the layers. \textbf{c} Absorption images of singly- and double-occupied lattices sites corresponding to the lattice configurations shown in b). The balanced bilayer configuration (top row) shows equal densities for both planes with a negligible contribution from double-occupied sites. The imbalanced bilayer shows the preferred population of the upper plane with mostly doubly-occupied lattice sites.}
	\label{Fig:SDInBothLayers}
\end{figure*}

In this work, we experimentally study the bilayer Hubbard model, see Figure 1a,  and examine the effect of different (local) chemical potentials in each of the layers on the thermodynamics. By adding a potential offset to one layer, we tune the chemical potential difference between the two layers. Therefore, one layer can be considered a  reservoir of energy and particles, and the tunnel coupling between the layers provides a bi-directional exchange between the two-dimensional planes.  We measure the equation of state $n(\mu)$ in each of the coupled layers by spatially-resolved detection of singly- and doubly-occupied sites. We observe, how different tunnel couplings and different values of the local chemical potential modify the equation of state and the derived thermodynamic potentials, such as the compressibility and the pressure. Additionally, we infer thermodynamic number fluctuations from the measured equation of state by employing the fluctuation-dissipation theorem and compare these thermodynamic observables to local density fluctuations. Here, we find that non-local density correlations are present in the system and increase with increasing $t_\perp /t$.

We create the bilayer Hubbard system following our previous implementation  \cite{gall_competing_2021}. In short, we load a quantum degenerate, spin-balanced mixture of the two lowest hyperfine states of the isotope $^{40}$K into an anisotropic optical lattice potential. In the vertical $z$--direction, the optical lattice potential composes of two standing waves with wavelength 1064\,nm and 532\,nm, respectively, with an adjustable relative phase and power, see Figure 1b. For the experiments conducted in this work, we choose a potential depth of 120\,$E_{rec}$ for the infrared lattice and either 15\,$E_{rec}$ or 25\,$E_{rec}$ for the green lattice. $E_{rec}$ denotes the recoil energy of the respective laser frequency. In the horizontal $xy$--plane, there are two, nearly orthogonal optical lattices formed by standing wave laser field of wavelength 1064\,nm with a potential depth of 6\,$E_{rec}$, corresponding to an in-plane tunneling rate of $t=h\times 174$\,Hz.  We utilize a digital micromirror device to compensate the confining potential created by the Gaussian envelope of the laser beams forming the optical lattice and use only the central part of the images for data analysis. The interaction strength between the atoms is determined by controlling the $s$--wave scattering length near a Feshbach resonance located at 202\,G, and for the experiments described here, we choose $U/t\simeq 8$. The temperature is $k_BT/t=2.2$.

In Figure 1c, we show  the principle of the measurement. We perform absorption imaging combined with spatial tomography in a magnetic field gradient  \cite{drewes_antiferromagnetic_2017,Wurz2018} in order to measure the density of singly-occupied sites $n^{(L)}_{i,\text{S}}=\braket{\hat{n}^{(L)}_{i,\uparrow}-\hat{n}^{(L)}_{i,\uparrow}\hat{n}^{(L)}_{i,\downarrow}}$  and the density  of doubly-occupied sites $n^{(L)}_{i,\text{D}}=\braket{\hat{n}^{(L)}_{i,\uparrow} \hat{n}^{(L)}_{i,\downarrow}}$ in either of the two layers $L=1,2$ for all lattice sites $i$ in the same experimental realisation. Additionally, we adjust the potential energy offset $\Delta$ between the two layers by adjusting the relative phase of the two optical lattices forming the superlattice, which leads to a differential chemical potential $\mu$ between the planes. 
If the detuning is chosen $\Delta = 0$, we obtain a symmetric configuration with both layers equally filled at approximately half filling. The repulsive interactions between the fermions induce a Mott insulator and we observe mostly singly-occupied lattice sites. If we induce a potential offset $\Delta$ between the wells, we realize an asymmetric superlattice configuration. For large values of $\Delta>U$  almost all atoms are located in the layer with the lower potential forming doubly-occupied lattice sites, and the second layer is empty.



Using the measured densities, we determine the equation of state $n^{(L)}_i\left(\mu_i^{(L)}\right)$ of the bilayer Hubbard model. Here, $n^{(L)}_i$  denotes the total  density per spin state $n^{(L)}_i  = n^{(L)}_{i,\text{S}} + n^{(L)}_{i,\text{D}}$, which is equal to the filling of the  lattice. For a balanced superlattice with $\Delta = 0$ the chemical potential is approximately $\mu=U/2$ in each layer, and we produce two equally-filled layers with a filling of up to 45\%, primarily limited by temperature \cite{gall_competing_2021}. In Figure 2a, we show the equation of state. The data are taken for both layers individually and for a given potential energy offset $\Delta$ they exhibit opposite signs of the chemcial potential.  We compare our experimental data to Determinant Quantum Monte Carlo (DQMC) simulations of a bilayer system with $6 \times 6\times 2$ lattice sites adapted from \cite{Varney2009} and find very good agreement.  The tunnel coupling between the layers induces an exchange of particles between the layers and hence affects the density fluctuations. Indications of this effect are the changes in the singles (Figure 2b) and doubles (Figure 2c) densities for different values of $t_\perp$ for the same chemical potential. Generally, a high density of singly-occupied sites and a low density of doubly-occupied sites is a signature of a Mott insulator in the individual layers. By increasing the inter-layer tunneling amplitude $t_{\perp}$, quantum disorder increases \cite{scalettar_magnetic_1994,kancharla_band_2007} and density fluctuations are energetically favourable. As a consequence, the Mott insulator melts, which we observe as a decrease of single-occupied lattice sites.

\begin{figure}
	\centering
	\includegraphics[width=0.5\textwidth]{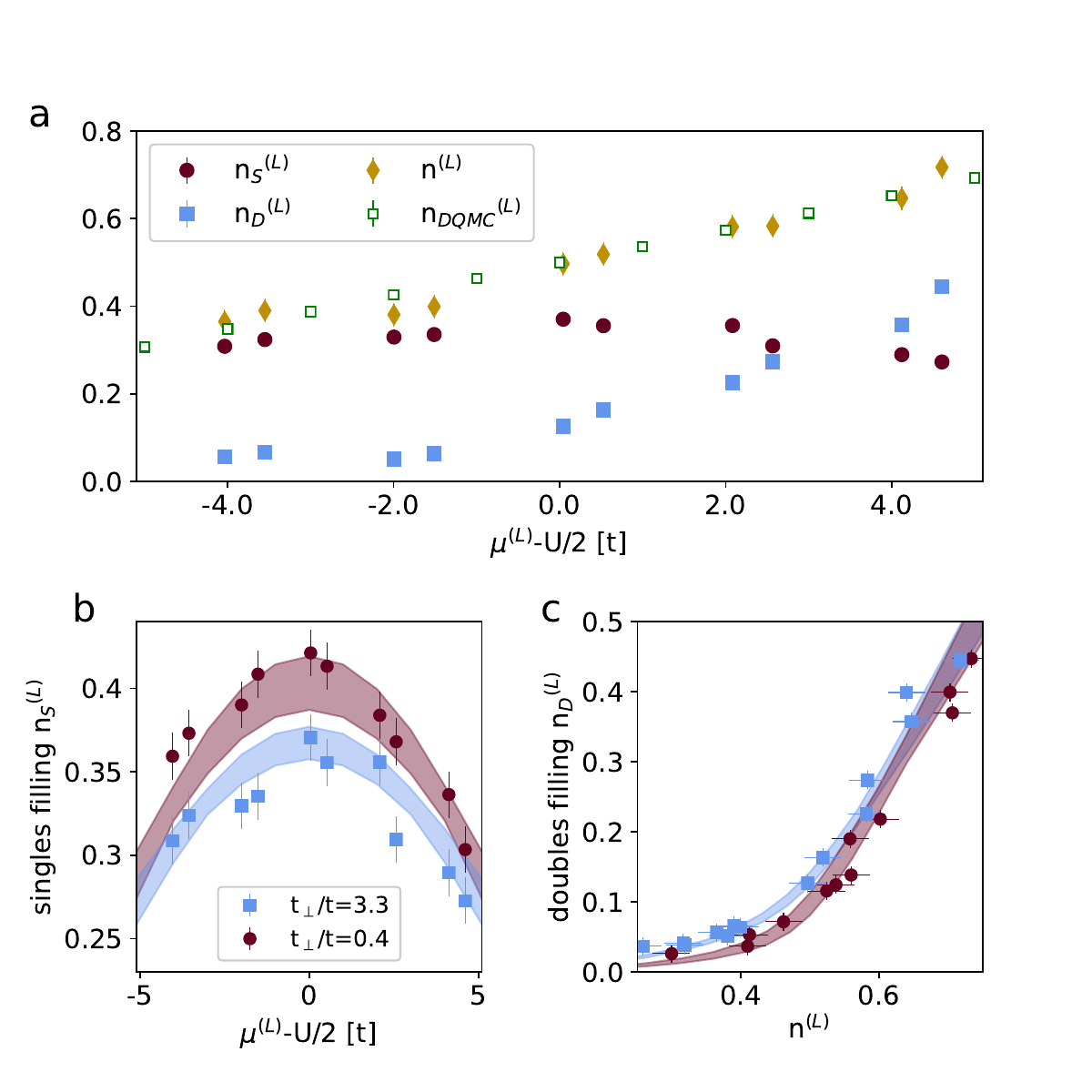}
	\caption{\textbf{a} Equation of state in the bilayer Hubbard model exemplarily for the ratio $t_\perp /t=3.3$. The chemical potential is tuned by varying the optical superlattice phase at a constant value of $U$.  \textbf{b} Singles density  as a function of the chemical potential for $t_\perp /t=0.4$ and $3.3$. \textbf{c} Doubles density as a function of the total filling for $t_\perp /t=0.4$ and $3.3$. The shaded regions in b and c are numerical simulations using DQMC and the width of the shaded region reflects the uncertainty of the entropy in the range $s =(1.28-1.31)k_\text{B}$.}
	\label{Fig:EoS}
\end{figure}

Next, we investigate the thermodynamic properties as a function of the tunnel coupling $t_\perp/t$ between the layers. To this end, we deduce the compressibility (Figure 3a) and the pressure (Figure 3b) from the equation of state. We define the iso-thermal compressibility as $\kappa^{(L)} = \partial n^{(L)}/ \partial \mu^{(L)}$, and for the numerical differentiation, we first interpolate $n^{(L)}$ and then differentiate this function with respect to $\mu^{(L)}$ \cite{Ku2012}. At around half-filling, $\kappa$ exhibits a suppression, which originates from the fact that it costs energy to produce double occupancies when compressing the repulsively interacting system. Overall, we find that the compressibility displays a fair agreement with the numerical simulations, however, the location of the minimum deviates from half-filling, which, at least in part, could be caused by the interpolation function used prior to  the numerical differentiation.


The pressure $p(\mu,T)$  for constant temperature is defined as
\begin{align}
	p(\mu,T) = \int _{-\infty} ^\mu n(\mu ' ,T)\,\text{d}\mu '.
\end{align}
For the numerical integration, we use the same interpolating function for $n^{(L)}$ as before. We observe an increase of the pressure with increasing total filling and, as theoretically expected, an almost linear dependence on the ratio $t_\perp/t$.

\begin{figure}
	\centering
	\includegraphics[width=0.5\textwidth]{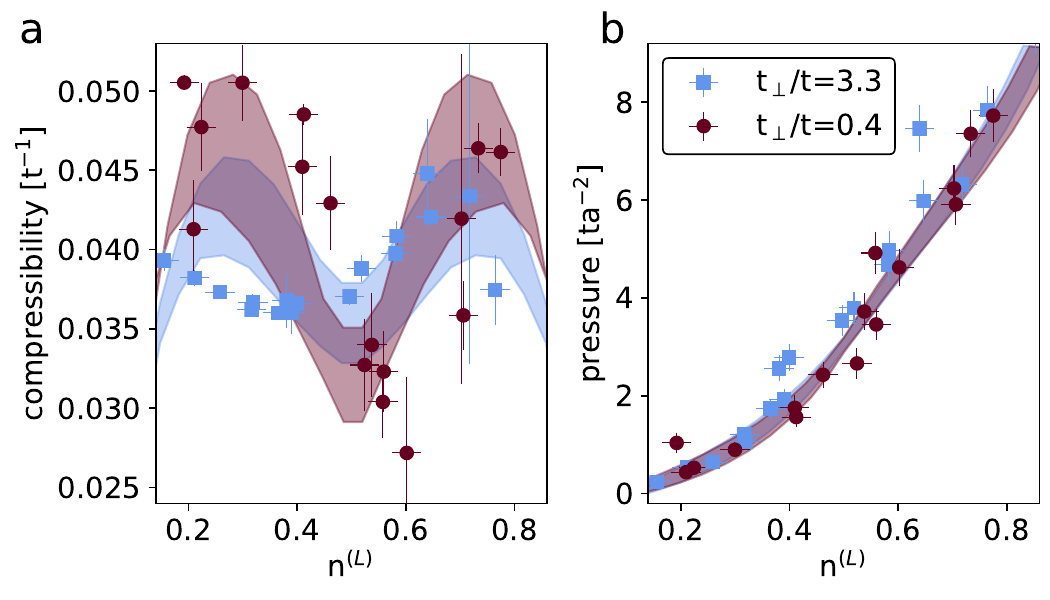}
	\caption{\textbf{a} Compressibility and \textbf{b} pressure as a function of the filling $n^{(L)}$ for the two ratios $t_\perp /t=0.4$ and $t_\perp /t=3.3$. The shaded regions in b and c are numerical simulations using DQMC and the width of the shaded region reflects the uncertainty of the entropy in the range $s =(1.28...1.31)k_\text{B}$. } 
	\label{Fig:CompPress}
\end{figure}


Thermal excitations and quantum tunnelling induce fluctuations of the density and from now on, we focus on the total density $\tilde{n}=n^{(L)}_\uparrow+n^{(L)}_\downarrow$. These fluctuations are characteristic of the quantum state of the system as they depend, for example, on the ratio of kinetic and interaction energy. Moreover, in the bilayer Hubbard system also the inter-layer coupling $t_\perp/t$ plays an important role, which we investigate by measuring density fluctuations for different interlayer coupling strength. 

Density fluctuations occur on all length scales -- from on-site to globally. Firstly, we study the on-site density fluctuations $\delta \tilde{n}_i^2 = \langle \hat{\tilde{n}}^2_i \rangle - \langle \hat{\tilde{n}}_i \rangle^2$ where $i$ denotes the lattice site index.  For a spin-balanced system, and by employing fermionic anti-commutation rules, we link the density fluctuations to the  measured singles and doubles densities
\begin{align}
	\delta \tilde{n}_i ^2 = 2\langle \hat{n}^{(L)}_{\downarrow, i} \rangle - 4\langle \hat{n}^{(L)}_{\downarrow, i} \rangle ^2 + 2\langle \hat{n}^{(L)}_{\downarrow, i}\hat{n}^{(L)}_{\uparrow, i} \rangle.
	\label{Eq:OnSiteFluc}
\end{align}
In Figure 4a, we show the measured density fluctuations as a function of filling for two different inter-layer coupling strengths  $t_\perp /t$ and compare to DQMC calculations. Generally, the on-site density fluctuations decrease with increasing filling since fermionic quantum statistics and repulsive interactions disfavour density fluctuations and the relevance of both increases with increasing filling. For the interaction strength of $U/t=8$ chosen here, the density fluctuations are in between the ideal Fermi gas limit $\delta n^2/n = 1-n/2$ (dashed line) and the infinite repulsion limit $\delta n^2/n = 1-n$ for $n \leq 0.5$ and $\delta n^2/n = 3 -n-2/n$  for $n > 0.5$ (solid line). In our work, we are specifically concerned with the role of the interlayer tunneling: a higher inter-layer tunneling rate increases the density fluctuations on a given lattice site since -- in addition to fluctuating within the layer -- the fermions can more easily tunnel into the adjacent layer.

Thermodynamic density fluctuations at a wave vector $q$ are quantified by the static structure  factor $S(q)$, which is the Fourier transform of the density-density correlation function. We employ the fluctuation-dissipation theorem \cite{Zhou2011,Kubo1966} to determine the thermodynamic density fluctuations at wave vector $q=0$  from the isothermal compressibility \cite{Klawunn2011, drewes_thermodynamics_2016} 
\begin{align}
	\frac{\kappa\, k_{\text{B}}T}{\tilde{n}} = S(q=0).
	\label{Eq:FlucDissTheo}
\end{align}
In Figure 4b we show our measurements of the static structure factor $S(q=0)$ for different fillings and inter-layer coupling strengths. The thermodynamic fluctuations are well below one, $S(q=0)<1$, as expected for the non-classical behaviour of a degenerate Fermi gas, and they decrease with increasing filling. For less than half filling the structure factor of the strongly-coupled layers ($t_\perp/t=3.3$) is below the structure factor of the weakly-coupled layers ($t_\perp/t=0.4$) whereas for filling greater than 0.5, the results are very similar.

\begin{figure}
	\centering
	\includegraphics[width=0.5\textwidth]{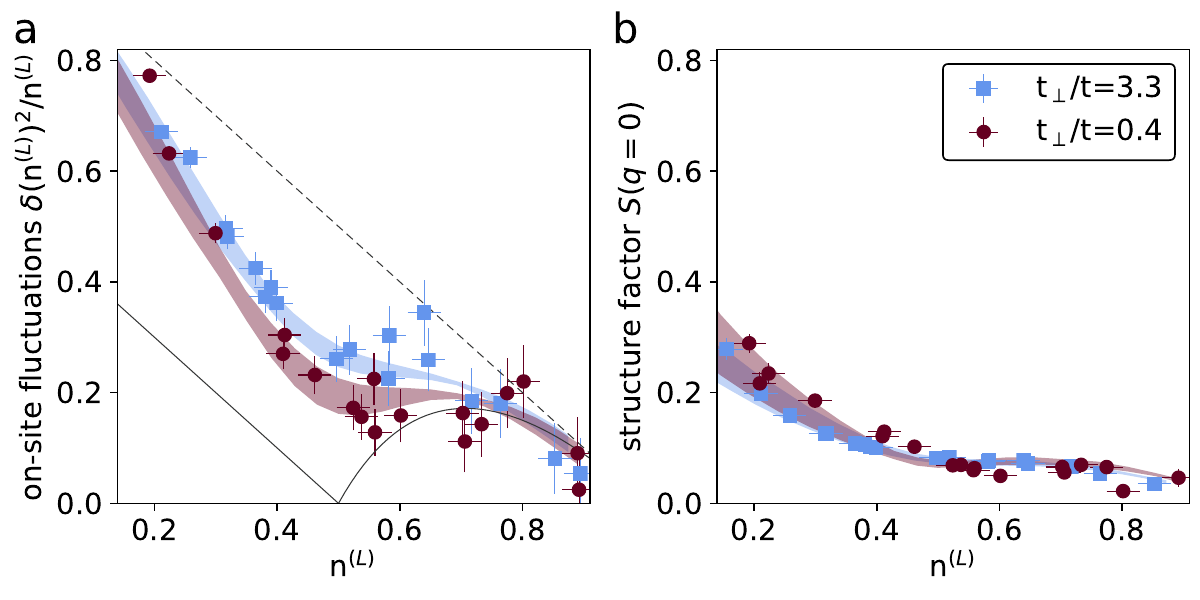}
	\caption{\textbf{a} On-site density fluctuations $\delta n_i^2/n_i$ as a function of the filling $n_i$ for the two ratios $t_\perp /t=0.4$ and $3.3$. \textbf{b} Static density structure factor $S(q=0)$ as a function of  $n_i$ for $t_\perp /t=0.4$ and $3.3$. The structure factor lies well below the measured on-site fluctuations, which indicates long-range density correlations, which increase with increasing ratio $t_\perp /t$. The shaded regions in b and c are numerical simulations using DQMC and the width of the shaded region reflects the uncertainty of the entropy in the range $s =(1.28-1.31)k_\text{B}$. The dashed (solid) lines shows the prediction for the non-interacting (infinite repulsively interacting) cases.  } 
	\label{Fig:StrFac_OnsiteFluc}
\end{figure}

We now compare the local with the thermodynamic fluctuations. We decompose the total density fluctuations into their on-site and off-site contributions \cite{Duchon2012,Fang2011}
\begin{align}
	\kappa = \frac{1}{a^2 k_\text{B} T} \left[ \delta \tilde{n}^2 + \sum _{j\neq i} (\langle \hat{\tilde{n}}_i\hat{\tilde{n}}_j\rangle - \langle \hat{\tilde{n}}_i \rangle\langle \hat{\tilde{n}}_j\rangle )\right].
\end{align}
Here, $a$ denotes the lattice constant. In a perfectly localised state, the non-local correlations are absent since $\langle \hat{\tilde{n}}_i\hat{\tilde{n}}_j\rangle = \langle \hat{\tilde{n}}_i \rangle\langle \hat{\tilde{n}}_j\rangle$. However, in a delocalised state, the non-local correlations are not negligible. Comparing Figures 4a and 4b shows that the difference between the thermodynamic and local density fluctuations increases for higher ratios $t_\perp /t$.  From this, we conclude that non-local density-density correlations are more pronounced for higher inter-layer tunneling $t_\perp /t$. This supports the qualitative concept that at half-filling for low inter-layer tunneling there are two weakly-coupled Mott insulators in each layer, whereas for strong inter-layer coupling singlets, which are delocalized across the bonds between the layers, dominate the physics.

In conclusion, we have investigated the thermodynamic properties of the bilayer Hubbard model and studied their dependence on the inter-layer tunneling. We find that a larger inter-layer tunneling decreases the compressibility and induces more non-local density correlations, except near half-filling where the trend is opposite because the singlets are more compressible than the two Mott insulators.

We thank C.F. Chan for contributions to the software code. This work has been supported by Deutsche Forschungsgemeinschaft through the Cluster of Excellence Matter and Light for Quantum Computing (ML4Q) EXC 2004/1–390534769 and SFB/TR 185 (project B4).

\end{document}